\documentclass[useAMS, usenatbib]{mn2e}

\usepackage[utf8]{inputenc}
\usepackage[T1]{fontenc}
\usepackage{aecompl}
\usepackage[english]{babel}

\usepackage[fleqn]{amsmath}
\usepackage{mathptmx} 
\usepackage[breaklinks,bookmarks,bookmarksnumbered]{hyperref}
\usepackage{graphicx}	

\newcommand{\mt}{}
\newcommand{\abs}[1]{\left|#1\right|}
\newcommand{\lambdat}{\Lambda_{\textrm{T}}}
\newcommand{\Ms}{M_{\textrm{s}}}
\newcommand{\Mp}{M_{\textrm{p}}}
\newcommand{\Mext}{\dot{M}_{\textrm{ext}}}

\renewcommand{\d}{\textrm{d}}

\newcommand{\uyr}{\mathrm{\,yr}}

\newcommand{\Msun}{\,M_{\odot}}

\usepackage{aas_macros}
\title[]{Estimating the fossil disc mass during supermassive black hole mergers: the importance of torque implementation}
\author[M. Tazzari and G. Lodato]{M. Tazzari$^{1,2,3}$\thanks{E-mail: \href{mailto:mtazzari@eso.org}{mtazzari@eso.org}.} and G. Lodato$^{3}$\\
$^{1}$European Southern Observatory (ESO), Karl-Schwarzschild-Str. 2, D-85748 Garching, Germany \\
${^{2}}$Excellence Cluster Universe, Boltzmannstr. 2, D-85748 Garching, Germany\\
$^{3}$Dipartimento di Fisica, Universit\`{a} degli Studi di Milano, Via Celoria 16, Milano, I-20133, Italy\\
}

\begin{document}
\date{}
\pagerange{\pageref{firstpage}--\pageref{lastpage}} \pubyear{2002}
\maketitle
\label{firstpage}

\begin{abstract}
In this paper, we revisit the issue of estimating the ``fossil'' disc
mass in the circumprimary disc, during the merger of a supermassive
black hole binary. As the binary orbital decay speeds up due to the
emission of gravitational waves, the gas in the circumprimary disc
might be forced to accrete rapidly and could in principle provide a
significant electromagnetic counterpart to the \mt{gravitational wave
emission}. Since the luminosity of such flare is proportional to the
gaseous mass in the circumprimary disc, estimating such mass
accurately is important. Previous investigations of this issue have
produced contradictory results, with some authors estimating
super-Eddington flares and large disc mass, while others suggesting
that the ``fossil'' disc mass is very low, even less than a Jupiter
mass. Here, we perform simple 1D calculations to show that such very
low estimates of the disc mass are an artifact of the specific
implementation of the tidal torque in 1D models. In particular, for
moderate mass ratios of the binary, the usual formula for the torque
used in 1D models significantly overestimates the width of the gap
induced by the secondary and this artificially leads to a very small
leftover circumprimary disc. Using a modified torque, calibrated to
reproduce the correct gap width as estimated by 3D models, leads to
fossil disc masses of the order of one solar mass. The rapid accretion
of the whole circumprimary disc would produce peak luminosities of the
order of 1-20 times the Eddington luminosity. Even if a significant
fraction of the gas escapes accretion by flowing out the secondary
orbit during the merger (an effect not included in our calculations),
we would still predict close to Eddington luminosities that might be
easily detected.
\end{abstract}

\begin{keywords}
accretion, accretion discs -- black hole physics -- hydrodynamics -- gravitational waves -- galaxies: formation
\end{keywords}

\section{Introduction}
The merger of two supermassive black holes (SMBH) is estimated to be one of the most intense events of gravitational wave (GW) emission and the ability to detect one such event is of paramount importance not only to test General Relativity directly but also to provide constraints on galaxy formation. 

There is now strong evidence for the presence of SMBHs with masses between ${10^{6}}$ and ${10^{9}\Msun}$ in the nuclei of the local Universe galaxies \citep{1995ARA&amp;A..33..581K, 1998AJ....115.2285M, 2005SSRv..116..523F}. In a hierarchical scenario for the galaxy evolution, nearby galaxies are the outcome of several mergers between progenitors of smaller mass. If each of the progenitors of a galaxy merger contains a SMBH in its centre the resulting galaxy will naturally host a pair of SMBHs. 

It is expected that dynamical friction between the binary and the crowded field of stars in which they are embedded can remove angular momentum from the binary and reduce its orbital separation down to parsec scales \citep{1980Natur.287..307B}, at which the process is expected to stall due to the depletion of the loss cone \citep{2001ApJ...563...34M}. Galaxy scale 3D hydrodynamical simulations \citep{2005ApJ...630..152E,2009MNRAS.393.1423C,2006MNRAS.372..869D} have shown that further shrinkage of the binary due to angular momentum loss to the gaseous background can lead to orbital separations of the order of 0.1 pc. At separations of the order of 0.001 pc, the main driver of orbital decay is the emission of GW. The evolution of the system between 0.1 and 0.001 pc is much harder to study. 3D and 2D simulations are appropriate to study the dynamics of binary-disc interaction on short timescales, and are thus well suited to study the final stages of the orbital decay, before and during the GW-driven phase \citep{2013MNRAS.436.2997D,2014ApJ...783..134F}. The overall evolution of the system during the long disc-driven decay is best followed using simple 1D diffusion models for the disc \citep{2002ApJ...567L...9A,2009MNRAS.398.1392L,2010MNRAS.407.2007C,2013ApJ...774..144R,2009ApJ...700.1952H, 2012MNRAS.427.2680K}.

In this paper, we want to estimate the mass left over in the circumprimary disc at decoupling, i.e. when the orbital decay due to GW emission becomes dominant. This quantity is the result of the long term evolution of the system from large distances ($\sim 0.1$ pc) down to the decoupling region. For our purposes, thus, it is most appropriate to study the problem within a 1D diffusion model. 

\begin{figure}
\center
\includegraphics[width=0.8\columnwidth]{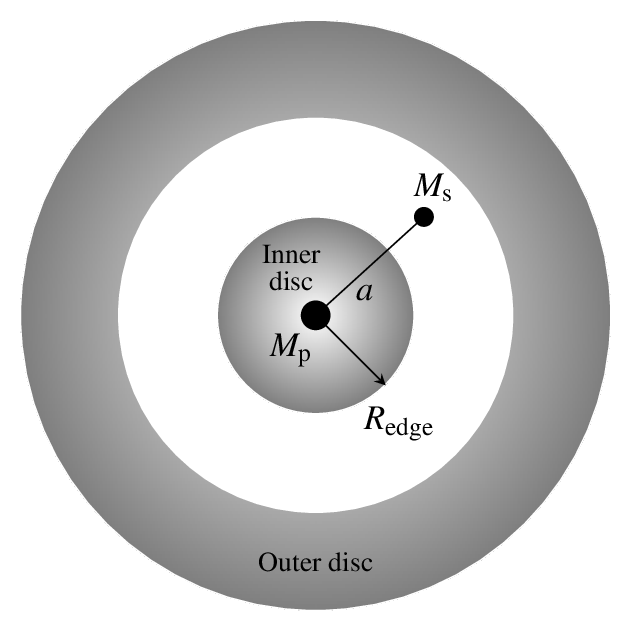}
\caption{Sketch of the binary system. The accretion disc, that orbits around the \textit{primary} BH of mass ${\Mp}$ is divided into two regions by the presence of the secondary BH of mass ${\Ms}$ that orbits in Keplerian motion at a radius ${a}$: a circumprimary disc (\textit{inner disc}) and a circumbinary disc (\textit{outer disc}). The inner disc is truncated at ${R_{\mathrm{edge}}}$ whose size depends on the intensity of the tidal torques.}
\label{fig.sketch} 
\end{figure}

In previous studies several configurations for the disc-binary system have been investigated. \citet{2002ApJ...567L...9A} assumed that the binary finds itself embedded in a gaseous disc that has already settled down to a steady-state disc. \citet{2009MNRAS.398.1392L} investigated a configuration   in which the binary interacts \mt{with a finite amount of gas that is brought in the binary vicinity as a consequence of the same accretion episode that has formed the binary itself}. \citet{2010MNRAS.407.2007C} considered the evolution of a binary embedded in a steady-state disc with a large mass inflow coming from the disc exterior. In all these configurations we can describe the system as sketched in Figure \ref{fig.sketch}: the accretion disc is rotating around a central (or \textit{primary}) BH of mass ${\Mp}$, and the \textit{secondary} BH of mass ${\Ms}$ is in Keplerian orbit at a radius ${a}$ around the primary. For simplicity, we assume that the primary BH is located at the centre of the accretion disc (thus neglecting the displacement between the primary BH and the centre of mass of the binary) and that the disc is  coplanar to the binary orbit \citep{1999MNRAS.307...79I}. The disc and the binary exchange angular momentum through tidal interaction in a planet-like dynamics: a secondary BH of non negligible mass opens up a gap in the disc at the locations where the tidal torques equal the viscous torques, with the gap following the secondary as it migrates (\textit{disc-driven} phase). A low-mass secondary BH barely perturbs the disc and thus migrates on a viscous timescale ${t_{\nu}=R^{2}/\nu}$, where ${R}$ is the radial cylindrical coordinate and ${\nu}$ is the disc viscosity. Conversely, a secondary BH of mass comparable to the disc is able to heavily perturb the disc by opening a wide gap but migrates more slowly.
Neglecting accretion onto the secondary BH, the disc results thus divided into two well defined regions: a \textit{circumprimary} (or \textit{inner}) disc orbiting around ${\Mp}$ and a \textit{circumbinary} (or \textit{outer}) disc orbiting around the whole binary. 

The investigations by \citet{2009MNRAS.398.1392L} and \citet{2010MNRAS.407.2007C} lead to apparently conflicting results. Indeed, \citet{2010MNRAS.407.2007C} estimate that the fossil disc mass is extremely low, of the order of $10^{-6}-10^{-3}\Msun$, depending on the parameters. On the other hand, \citet{2009MNRAS.398.1392L}, while not giving specific numbers for the inner disc mass, clearly show that the surface density at decoupling can be much higher than estimated by \citet{2010MNRAS.407.2007C}. The origin of this discrepancy is unclear, and it could be due to the different setup and initial conditions used in the two studies. In this paper, we aim to resolve the issue by re-evaluating the inner disc mass, using a disc configuration close to the one used by \citet{2010MNRAS.407.2007C}. 

As the binary decays further, the angular momentum loss due to the emission of GW rapidly increases and eventually takes over (\textit{GW-driven} phase). We define the \textit{decoupling} time ${t_{\mathrm{dec}}}$ as the time when the binary merger timescale becomes shorter than the viscous timescale at the inner edge of the outer disc, i.e. when the outer disc is not able to follow viscously the accelerating secondary and thus decouples from the evolution of the binary \citep[for a detailed study on the decoupling process see][]{2014ApJ...783..134F}. Assuming for simplicity that the inner disc can drain only by accretion onto the primary BH, in the late GW-driven phase the inner disc is forced to accrete by the plunging secondary BH that migrates much more rapidly. The extremely rapid accretion of the whole inner disc (whose mass is substantially frozen during the GW-driven phase) might lead to a super-Eddington flare. Note that \citet{2012MNRAS.423L..65B}, using 2D simulations, show that such rapid accretion can be reduced due to the funneling of the inner gas out of the secondary orbit during the GW-driven phase, estimating that up to 80 percent of the inner disc mass can escape accretion in this way. Such results, however, need to be confirmed in 3D simulations and using a wider range of parameters, especially for the disc aspect ratio (which \citealt{2012MNRAS.423L..65B} fix to a relatively large value).  

The paper is organized as follows. In Section~\ref{sec:physical.model} we illustrate the physical model that we adopt to describe the disc-binary system with a detailed discussion of the torque implementation. In Section~\ref{sec:results} we describe the results of a set of simulations with different parameters and the overall evolution of a fiducial run. In Section~\ref{sec:discussion} we discuss our results in comparison with those in literature and we propose an analytic argument to explain the discrepancy in the estimate of the inner disc mass. In Section~\ref{sec:conclusions} we present our conclusions and the outlook for this work.

\section{Physical model}
\label{sec:physical.model}
In order to describe the coupled evolution of the disc-binary system, we have implemented a time-dependent 1D model that solves for the disc surface density ${\Sigma(R,t)}$ and for the binary separation ${a(t)}$ in a self-consistent way. 

\subsection{Disc dynamics}
The evolution of the disc surface density in presence of the secondary BH is described by the classical hydrodynamical equations of accretion theory \citep{1974MNRAS.168..603L,1981ARA&amp;A..19..137P}: the continuity equation
\begin{equation}
\label{eq.continuity}
\frac{\partial \Sigma}{\partial t} + \frac{1}{R}\frac{\partial}{\partial R}\left(\Sigma R v_{R}\right)=0\,,
\end{equation}
and the angular momentum conservation
\begin{equation}
\label{eq.angular.momentum.conservation}
\frac{\partial}{\partial t}\left(\Sigma R v_{\varphi}\right)+\frac{1}{R}\frac{\partial}{\partial R}\left(\Sigma R v_{\varphi} R v_{R}\right)
=
\frac{1}{2\pi R}\frac{\partial G_{\nu}}{\partial R} + \Lambda_{\mathrm{T}}\Sigma
\,,
\end{equation}
where ${\vec v = (v_{R}, v_{\varphi}, v_{Z})}$ is the fluid velocity, ${G_{\nu}=2\pi \nu \Sigma R^{3} \Omega'}$ is the usual viscous torque and ${\Lambda_{\mathrm{T}}}$ is the specific tidal torque exerted by the binary on the disc. Using \eqref{eq.continuity} and \eqref{eq.angular.momentum.conservation} we obtain an expression for ${v_{R}}$ and substituting it back in \eqref{eq.angular.momentum.conservation} we obtain the time evolution equation:
\begin{equation}
\label{eq.disc.evolution}
\frac{\partial \Sigma}{\partial t} 
=\frac{3}{R}\frac{\partial}{\partial R}\left[R^{1/2} \frac{\partial}{\partial R}\left(R^{1/2}\nu \Sigma \right) \right] - \frac{2}{R}\frac{\partial}{\partial R}\left[ \frac{\lambdat \Sigma}{\Omega} \right]\,,
\end{equation}
where ${\nu(R)}$ is the disc viscosity and ${\Omega(R)=\sqrt{G\Mp/R^{3}}}$ is the disc angular velocity.

Making use of the \textit{impulse approximation} \citep{1979MNRAS.188..191L, 1987gady.book.....B} we can compute a simplified expression for ${\lambdat(R)}$ that has been widely used in literature \citep{1979MNRAS.186..799L,2002ApJ...567L...9A,2004MNRAS.353..841L,2009MNRAS.398.1392L}:
\begin{align}
\label{eq.tidal.torque.original1}
\lambdat &=- \frac{f}{2} q^{2}\Omega^{2}R^{2}\biggl(\frac{R}{\Delta} \biggr)^{4} \quad R < a \,, \\
\label{eq.tidal.torque.original2}
\lambdat &= \frac{f}{2} q^{2}\Omega^{2}R^{2}\biggl(\frac{a}{\Delta} \biggr)^{4} \quad R> a \,,
\end{align}
where ${f}$ is a normalization factor and ${\Delta=R-a}$. The change of sign across ${R=a}$ in the tidal torque accounts for the behaviour of the secondary BH that removes angular momentum from the gas inside its orbit and adds angular momentum to the gas outside. The net effect is that the satellite repels the gas from its orbit, thus creating an annular \textit{gap} whose width depends on the intensity of the tidal torques with respect to the pressure and viscosity gradients that act to close the gap. 

Since the above expression diverges in ${R=a}$, following \citet{1986ApJ...309..846L} and \cite{1995MNRAS.277..758S} we smooth the torque by taking:
\begin{equation}
\label{eq.smoothing.secondary}
\Delta = \max \left\{H,R_{\mathrm{H}},\abs{R-a} \right\}\,,
\end{equation}
where 
${H}$ is the disc thickness and ${R_{\mathrm{H}}=a \left(q/3 \right)^{1/3}}$ is the Hill's sphere radius of the secondary BH. This smoothing prescription follows from the basic idea that the gap should be larger than both ${H}$ (otherwise it would be replenished by the pressure gradients in less than a dynamical timescale) and ${R_{\mathrm{H}}}$ (that should be depleted due to the accretion on the secondary BH)\footnote{We should note that apart from this smoothing prescription we have not taken into account the accretion on the secondary BH in the model. \mt{This is a good approximation for ${q=0.1}$ \citep{2014ApJ...783..134F}, but some accretion is expected for $q\ga 0.3$, which might slow down migration somewhat.}}. 

It is easy to see that the intensity of the tidal torque determines the gap width since the gap edges are the locations where the viscous torques, that try to refill the gap with a diffusive behaviour, are balanced by the tidal torques that act to keep the gap open. Analogously to ${\lambdat}$, we can define the specific viscous torque ${\Lambda_{\nu}}$ dividing the annular viscous torque ${G_{\nu}}$ by the mass of the annulus, namely:
\begin{equation}
\Lambda_{\nu}=\frac{G_{\nu}}{2\pi R \Sigma \Delta} =
\frac{\nu R^{2}}{\Delta}\Omega' = -\frac{3}{2}\frac{\Omega R}{\Delta}\nu\,,
\end{equation}
where the last equality holds for a Keplerian disc. The gap width is thus the value of ${\Delta}$ for which ${\Lambda_{\nu}=\lambdat}$. In Figure 2 we illustrate how, if the intensity of ${\lambdat}$ changes, then the location of the gap edges changes as well: for a given viscous torque intensity (that in the figure is assumed constant for simplicity) a stronger tidal torque truncates the inner disc at a smaller radius. Therefore, in order to implement a realistic tidal torque, the parameters describing the torque (${f}$ and the smoothing prescription) have to be chosen carefully, so as to reproduce the gap size obtained by more sophisticated 3D simulations. 
\begin{figure}
\includegraphics[scale=1]{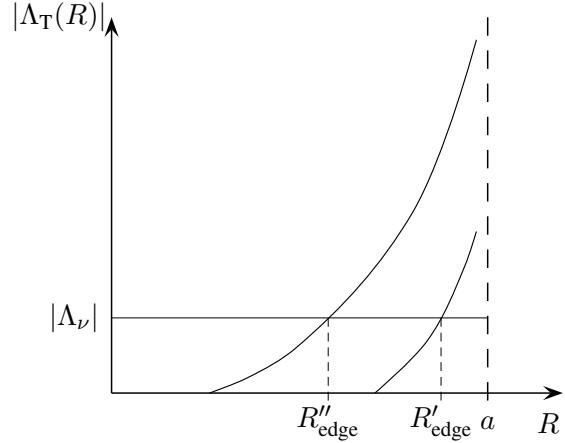}
\caption{Different torque intensity induce different gap sizes.  Two torques of different intensity are shown, the upper one being more intense than the lower one. The gap inner edge (${R_{\mathrm{edge}}}$) is located where the tidal torque ${\lambdat}$ equals the viscous torques ${\Lambda_{\nu}}$ (that here for simplicity has been fixed to be constant). It can be seen that, for a given ${\Lambda_{\nu}}$, the gap edge produced by the stronger torque is located at ${R_{\textrm{edge}}''}$, that is smaller (i.e. the gap is wider) than that produced by the weaker torque (${R_{\textrm{edge}}'}$). The same result applies for ${R>a}$.}
\label{fig.gap.edges}
\end{figure}

Since the size of the gap in the disc, as well as the migration rate of the secondary, depends on the intensity and the spatial distribution of the tidal torques, a correct smoothing of the tidal torques is necessary not only to avoid numerical issues, but mostly  to avoid an unphysical estimate of the interaction between the disc and the binary that can lead to a wrong estimate of the inner disc mass.

\subsection{Tidal torque smoothing}
\label{sec.tidal.orque}
As shown by \citet{1979ApJ...233..857G,1980ApJ...241..425G} the disc and the secondary interact by exchanging angular momentum at discrete \textit{resonant} locations, the so-called \textit{Lindblad Resonances}, where a characteristic frequency of the secondary matches an epicyclic frequency within the disc: at these locations the disc responds to the presence of the secondary through the excitation of density waves that carry energy and angular momentum in the disc and dissipate through shocks. 

In a Keplerian disc the Lindblad Resonances are located at radii:
\begin{equation}
R_{\textrm{L}}=a \left( 1\pm \frac{1}{m} \right)^{2/3} \qquad m=1,2,...
\end{equation}
which shows that for low ${m}$ the Lindblad resonances are far from the position of the secondary whereas for increasing ${m}$ they accumulate towards ${R=a}$. We find that the \mt{innermost Lindblad Resonance (hereafter, IMLR)} is placed at ${R_{\mathrm{IMLR}}=2^{-2/3}a\simeq 0.63 a\,}$, whereas the outermost one (OMLR) is placed at ${R_{\mathrm{OMLR}}=2^{2/3}a\simeq 1.59 a\,}$. Since the IMLR and the OMLR are the farthermost resonances from the satellite, they constitute the boundaries of the region that can effectively contribute to the tidal torque: outside this region, the disc can exert no torque on the satellite, nor the satellite can exert any torque on the disc. As noted in \citet{2009MNRAS.398.1392L}, since ${\lambdat}$ scales as ${q^{2}}$, for high ${q}$ it can occur that the analytical expressions \eqref{eq.tidal.torque.original1} and \eqref{eq.tidal.torque.original2} gives a non negligible contribution also outside  the region of LRs. In order to avoid this unphysical contribution, we need to smooth out the torque for ${R<R_{\mathrm{IMLR}}}$ and ${R>R_{\mathrm{OMLR}}}$. We can do this by fine-tuning the torque formulae \eqref{eq.tidal.torque.original1} and \eqref{eq.tidal.torque.original2} in two different ways, depending on the mass ratio of the binary.

For ${q\ll 1}$ the torque formulae \eqref{eq.tidal.torque.original1} and \eqref{eq.tidal.torque.original2} already gives no contribution outside the IMLR and the OMLR, and the gap size is determined by the parameter ${f}$. Therefore, in this case, it is sufficient to fine-tune the value of ${f}$ to reproduce realistic gap sizes. Following \citet{2002ApJ...567L...9A} that performed high-resolution 2D simulations using the ZEUS hydrodynamic code \citep{1992ApJS...80..753S}, we take ${f=10^{-2}}$ with which we reproduce gaps of approximately the correct size. 

For larger mass ratios, the torque is still significant outside the LRs region and thus we smooth it out with a gaussian cutoff by setting:
\begin{align}
\lambdat &= -\dfrac{f}{2} q^{2}\Omega^{2}R^{2}\biggl(\dfrac{R}{\Delta} \biggr)^{4}\exp\left[-\left(\dfrac{R-R_{\mathrm{IMLR}}}{w_{\mathrm{IMLR}}} \right)^{2} \right] \ R\leq R_{\mathrm{IMLR}} \\[1em]
\lambdat &= \quad\!\dfrac{f}{2} q^{2}\Omega^{2}R^{2}\biggl(\dfrac{a}{\Delta} \biggr)^{4} \exp\left[-\left(\dfrac{R-R_{\mathrm{OMLR}}}{w_{\mathrm{OMLR}}} \right)^{2} \right] \ R \geq R_{\mathrm{OMLR}}
\end{align}
where ${w_{\mathrm{IMLR}}}$ and ${w_{\mathrm{OMLR}}}$ determine the sharpness of the cutoff and for ${R_{\mathrm{IMLR}}< R < R_{\mathrm{OMLR}}}$ the formulae \eqref{eq.tidal.torque.original1} and \eqref{eq.tidal.torque.original2} still hold. We fine-tune the values of ${w_{\mathrm{IMLR}}}$ and ${w_{\mathrm{OMLR}}}$ so that the gap edges (measured after the secondary has opened a clear gap in the disc) are located in the same positions computed by \citet{1994ApJ...421..651A} with 3D SPH simulations. \citet{1994ApJ...421..651A} studied circumbinary discs around binaries with ${q=0.11}$ and ${q=0.43}$, on circular and eccentric orbits, and for discs with different Reynolds numbers. Furthermore, \citet{1994ApJ...421..651A} and \citet{1977MNRAS.181..441P} showed that for zero-eccentricity binaries the location of the gap edges does not depend on the value of the Shakura-Sunyaev viscosity parameter ${\alpha}$ (see Eq. \ref{eq.alpha.prescription}). In Table \ref{tab.smoothing.lengths} we report the values of ${w_{\mathrm{IMLR}}}$ and ${w_{\mathrm{OMLR}}}$ that we obtain for each mass ratio. Since the binary we are considering for our purposes is on a circular orbit and the values found for ${w_{\mathrm{IMLR}}}$ and ${w_{\mathrm{OMLR}}}$ weakly depend on the mass ratio, for our simulations we always adopt the values ${w_{\mathrm{IMLR}}=370H}$ and ${w_{\mathrm{OMLR}}=75H}$.

\begin{table}
\caption{Inner and outer \textit{gap edges} obtained by \citet{1994ApJ...421..651A} performing 3D SPH simulations, for ${\alpha=0}$. We report the widths ${w_{\mathrm{IMLR}}}$ and ${w_{\mathrm{OMLR}}}$ of the gaussian smoothing that is needed to reproduce the same gap edges.}
\begin{tabular}{@{}ccccc}
\hline
${q}$ & \multicolumn{2}{c}{Gap Edges} & \multicolumn{2}{c}{Gaussian widths} \\[0.2em]
 & Inner & Outer & ${w_{\mathrm{IMLR}}}$ & ${w_{\mathrm{OMLR}}}$ \\
\hline
0.11 & 0.46${a}$ & 1.8${a}$ & 370${H}$ & 75${H}$ \\
0.43 & 0.38${a}$ & 1.8${a}$ & 370${H}$ & 83${H}$ \\
\hline
\end{tabular}
\label{tab.smoothing.lengths}
\end{table}

\mt{Note that the formula used here for the local torque is clearly an approximation, used in our 1D model to reproduce the correct gap width and secondary migration rate. Recent simulations by \citet{2014ApJ...783..134F} allow to compute the local torque as a function of radius. The torque obtained in this case has a different functional form from the one used here. In particular, it has an oscillatory behaviour and does not vanish until 2-3a. However, for comparable values of $q$, Farris et al confirm the gap widths found by \citet{1994ApJ...421..651A}. The exact shape of the torque close to the secondary does not affect any of our results (indeed, in many cases one can treat the secondary, simply imposing an effective boundary condition, see \citealt{1995MNRAS.277..758S}), as long as the gap width and migration rate are reproduced correctly. If we had kept our analytical torque formula, allowing it to extend to 2-3a \citep{2014ApJ...783..134F}, we would have obtained an incorrect gap width. Hence we have preferred to be consistent with our approach and smooth the torque in order to reproduce reasonable gap widths.}

\subsection{Disc energetics}
\label{sec.disc.energetics}
In an accretion disc surrounding a SMBH binary we expect to find regions characterized by different regimes, depending on the distance from the central BH, the temperature, the disc thickness, etc. Following the approach adopted in \citet{2009MNRAS.398.1392L}, our code is capable of reproducing both radiation-pressure and gas-pressure dominated regimes (we expect the first to occur in the inner part of the disc and the latter in the outer ones), and also to take into account both Thomson scattering and free-free absorption as sources of opacity.

In order to compute the viscosity of the disc, we adopt the standard ${\alpha-}$prescription \citep{1973A&A....24..337S}:
\begin{equation}
\label{eq.alpha.prescription}
\nu = \alpha c_{\mathrm{s}} H\,,
\end{equation}
where ${c_{\mathrm{s}}}$ is the sound speed and ${\alpha}$ is a dimensionless number smaller than unity. In order to avoid viscous and thermal instabilities in the disc, following \citet{1974ApJ...187L...1L} we assume that the viscous stresses are proportional to the gas pressure (and not to the \textit{total} pressure, sum of gas and radiation contributions), and therefore ${c_{\mathrm{s}}=\sqrt{k_{\mathrm{B}}T_{\mathrm{c}}/\mu m_{\mathrm{p}}}}$, with ${k_{\mathrm{B}}}$ being the Boltzmann constant, ${T_{\mathrm{c}}}$ the midplane temperature, ${\mu=0.67}$ the mean molecular weight (ionized Hydrogen) and ${m_{\mathrm{p}}}$ the proton mass.  

The midplane temperature profile is computed by solving the vertical energy balance at each radius. First, we suppose that the energy dissipated by the viscous processes (released mainly in the disc midplane) is completely irradiated by the disc at its surface by imposing:
\begin{equation}
\label{vertical.energy.balance}
\frac{9}{8}\nu \Sigma \Omega^{2} = \sigma_{\mathrm{SB}} T_{\mathrm{s}}^{4}\,,
\end{equation}
where ${\sigma_{\mathrm{SB}}}$ is the Stefan-Boltzmann constant and ${T_{\mathrm{s}}}$ is the surface temperature. Then, we assume that the disc is optically thick to its own radiation, i.e. the optical depth ${\tau=\kappa \Sigma /2 \gg 1}$, where ${\kappa}$ is the opacity (given by the dominant source among Thomson scattering or free-free absorption). Finally, plugging the ${\alpha-}$prescription \eqref{eq.alpha.prescription} in Eq. \eqref{vertical.energy.balance} and using the fact that ${T_{\mathrm{s}}=4T_{\mathrm{c}}/3\tau^{1/4}}$, we find that the midplane temperature is given by:
\begin{equation}
\label{4.tcentral}
T_{c}=\left(\frac{9}{16}\frac{\alpha k_{\mathrm{B}}}{\mu m_{\mathrm{p}} \sigma_{\mathrm{SB}}}\kappa\Omega \Sigma^{2} \right)^{1/3}\,.
\end{equation}
Once computed the viscosity and the central temperature, the disc thickness can be calculated from the vertical hydrostatic equilibrium that results from the balance between gravity and the total pressure.
At each radius the structure of the disc is fully determined by ${\nu}$, ${T_{\mathrm{c}}}$ and ${H}$, and we use an iterative procedure that converges to consistent values for these three parameters. 

\mt{Note that \citet{2009MNRAS.398.1392L} have shown that, in order to simultaneously conserve energy and angular momentum in the 1D formulation that we adopt here, a tidal heating contribution should be added at the disc edge to account for the dissipation of non-axisymmetric structures induced by the tidal force. We neglect here this contribution, mostly to compare more directly to \citet{2010MNRAS.407.2007C} (who also do not include it). This tidal heating term could modify some of the results in limited portion of the parameter space (see \citealt{2012MNRAS.427.2680K}), when the gap is not fully open.}

\subsection{Binary dynamics}
To derive the equation for the binary evolution we simply impose that the angular momentum of the whole disc-binary system is conserved and we take into account the angular momentum lost due to the emission of GW. The binary separation thus evolves as follows:
\begin{equation}
\label{eq.binary.evolution}
\dot{a}=-\frac{4\pi}{a \Ms \,\Omega_{\mathrm{s}}}\int_{R_{\mathrm{in}}}^{R_{\mathrm{out}}}\lambdat \Sigma\, R \d R-\frac{8c}{5}q(1+q)\left(\frac{R_{\mathrm{S}}}{a} \right)^{3}
\end{equation}
where ${\Omega_{\mathrm{s}}=\sqrt{G\Mp/a^{3}}}$ is the angular velocity of the secondary and ${R_{\mathrm{S}}}$ is the Schwarzschild radius of the primary, ${R_{\mathrm{S}}=2G\Mp/c^{2}}$. The first term on the right hand side is the tidal back reaction of the disc on the binary and the second term is due to the GW emission \citep{1963PhRv..131..435P, 1964PhRv..136.1224P}. 

We note that while the direction of migration due to the tidal torque can change in time since it depends both on the intensity of the torque and on the surface density distribution, the emission of GW always acts to shrink the binary.

\section{Results}
\label{sec:results}
\subsection{Numerical code: setup and initial configuration}
We solve numerically the coupled equations \eqref{eq.disc.evolution} and \eqref{eq.binary.evolution} with a 1D finite-difference numerical scheme that allow us to compute the time evolution time of ${\Sigma(R,t)}$ and ${a(t)}$. We use a radial logarithmic grid of 200 points, with the inner radius equal to the innermost stable circular orbit (ISCO) around the central BH, i.e. ${R_{\mathrm{in}}=3R_{\mathrm{S}}}$, and the outer radius ${R_{\mathrm{out}}=10a_{0}}$ where ${a_{0}=10^{4}R_{\mathrm{S}}}$ is the initial position of the secondary BH. At ${R_{\mathrm{in}}}$ we apply a zero-torque boundary condition to implement the accretion on the central BH, whereas at ${R_{\mathrm{out}}}$ we can choose to set either a zero-torque boundary condition or an inflow of material ${\Mext}$, depending on the scenario we want to model.

In previous investigations, \citet{2002ApJ...567L...9A} initialized the disc with a steady-state disc corresponding to ${\Mext=1\Msun\uyr^{-1}}$, while during the simulation they assumed ${\Mext=0}$. With this choice, the secondary BH results parachuted inside a disc that has already settled down to a steady state and therefore the binary hardening is driven by the almost infinite reservoir of gas upstream and not by an inflow of material coming from ${R_{\mathrm{out}}}$. More recently, \citet{2009MNRAS.398.1392L} investigated a scenario in which the SMBH binary interacts with a finite amount of gas that evolves together with the binary and is far from a steady state: as initial configuration they assumed a constant surface density profile between ${0.8a_{0}}$ and ${2a_{0}}$ and ${\Mext=0}$ to simply model this scenario.
We performed several tests of our code using these two configurations and we are able to reproduce the results of \citet{2002ApJ...567L...9A} and \citet{2009MNRAS.398.1392L} to a high degree of accuracy. 

Despite being less realistic than the previous scenario, to compare our results to the work of \citet{2010MNRAS.407.2007C}, in all our simulations we choose initialize the disc to a steady-state disc of low mass. To do this, instead of adopting analytic solutions, we have obtained the steady state configurations by solving numerically with our code Eq. \eqref{eq.disc.evolution} without the tidal term on the rhs and applying a ${\Mext \neq 0}$. To probe the accuracy of our code, we successfully proved that the surface density profile of this initial steady state is in good agreement with the standard solutions of the Shakura-Sunyaev ${\alpha-}$disc. \mt{The initial accretion rate through the disc is thus constant and is set to a very low value, much smaller that the $\dot{M}_{\mathrm{ext}}$ that we then apply. As the simulation starts, the disc starts to fill up with mass and the accretion rate thus starts to slowly grow, as the disc tries to evolve toward the new configuration appropriate for the higher accretion rate}. At the initial time-step, we set the position of the secondary BH in ${R=a_{0}}$ and we let the system evolve: at the beginning, the secondary BH clears a gap in the disc and, once the tidal interaction becomes strong enough, it starts migrating.

\subsection{Resulting fossil disc mass}

How much mass is left in the inner disc at decoupling has been the subject of debate recently. \citet{2002ApJ...567L...9A} estimate a relatively large mass based on very idealized initial conditions. \citet{2009MNRAS.398.1392L}, using more realistic conditions found a relatively large mass, too, with super-Eddington accretion rates once the inner disc is rapidly accreted during the GW-driven binary decay. Conversely, \citet{2010MNRAS.407.2007C} found that the fossil disc mass is extremely low, being of the order of planetary masses. Thus, our main aim is to resolve the discrepancy and estimate what is the mass of the fossil disc at decoupling, comparing our results to those of \citet{2010MNRAS.407.2007C}. We thus perform a number of simulations, using the very same parameters of \citet{2010MNRAS.407.2007C}. Our code is similar to the one used by \citet{2010MNRAS.407.2007C} in many respects and we would thus expect to reproduce their results. 

In the simulations that we perform we are interested in computing the evolution of the accretion luminosity ${L_{\mathrm{acc}}}$ as the binary approaches to the merger and we do that by simply computing ${L_{\mathrm{acc}}=\eta \dot{M}c^{2}}$ where ${\eta=0.1}$ is the accretion efficiency and ${\dot{M}}$ is the instantaneous mass accretion rate onto the primary BH \mt{(measured as the mass per unit time flowing inside the ISCO, i.e. $R<{R_{{in}}}$)}. As described above, we call \textit{decoupling} radius the location where the merger time driven by the emission of GW becomes smaller than the merger time due to the tidal interaction with the disc. At that point, the binary decouples from the dynamics of the disc and evolves on a timescale that is too short for the disc to respond viscously \citep[but see][]{2014ApJ...783..134F}. During this phase, the mass of the inner disc is substantially frozen since the secondary forces it to accrete onto the primary at a rate much higher than the viscous one. During the evolution of the system, we compute ${L_{\mathrm{acc}}}$ from ${\dot{M}}$ at each timestep and we estimate the peak luminosity ${L_{\mathrm{peak}}=max(L_{\mathrm{acc}})}$ that is reached immediately before the merger. 
We can put a lower limit to ${L_{\mathrm{peak}}}$ estimating the average luminosity that is produced by the accretion of the inner disc in the time interval between the decoupling and the merger, namely
\begin{equation}
\label{eq.average.lacc}
L_{\mathrm{peak}}\geq \frac{M_{\mathrm{d,in}}(t_{\mathrm{dec}})}{t_{\mathrm{merger}}-t_{\mathrm{dec}}}\eta c^{2}\,,
\end{equation}
where ${M_{\mathrm{d,in}}(t)}$ is the time-dependent mass of the inner disc, defined as follows:
\begin{equation}
M_{\mathrm{d,in}}(t)=\int_{R_{\mathrm{in}}}^{R_{\mathrm{edge}}}2\pi R\, \Sigma(R, t) \d R\,,
\end{equation}
where ${R_{\mathrm{edge}}}$ is the outer edge of the inner disc, as shown in Figure \ref{fig.sketch}. Clearly, this estimate is valid if we assume that the whole inner disc is accreted during the merger \citep{2012MNRAS.423L..65B}.

We perform a series of simulations exploring different regions of the parameters space by varying the primary BH mass ${\Mp =10^{6},\ 10^{7},\ 10^{8} \Msun}$, the external accretion rate ${\dot{M}_{\mathrm{ext}}=0.1,\ 0.01 \Msun \uyr^{-1}}$, the mass ratio ${q=0.3,\ 0.1}$ and the Shakura-Sunyaev viscosity parameter ${\alpha=0.1,\ 0.01}$, whereas we keep fixed the initial separation ${a_{0}=10^{4}R_{\mathrm{S}}}$ and the initial disc mass ${M_{\mathrm{d}}=10^{3}\Msun}$. 

The results of the simulations are shown in Table \ref{table.results}. We report the inner disc mass at decoupling ${M_{\mathrm{d,in}}}$ and the peak value of the accretion luminosity ${L_{\mathrm{peak}}}$ just prior to merger. In almost all the cases we obtain peak luminosities that clearly exceed the Eddington luminosity and in four cases the flare is super-Eddington by one order of magnitude. It is remarkable that even though ${M_{\mathrm{d,in}}}$ is relatively small (in many cases a few percent of one solar mass), its forced accretion during the final inspiral of the secondary results in an extremely luminous flare. 

The last two columns in Table 2 show the fossil disc mass and peak luminosity as obtained by \citet{2010MNRAS.407.2007C}. There is a striking difference in the estimated disc mass by up to three orders of magnitude with respect to ours. This systematic difference is found for all parameters used and results also in a significant difference in peak luminosity. We discuss the origin of this discrepancy in section \ref{sec:discussion}, after describing the dynamical evolution of a fiducial run. 

\mt{Note that while \citet{2010MNRAS.407.2007C} obtain an almost linear relation between the fossil disc mass ${M_{\textrm{d,in}}}$ and the external accretion rate $\dot{M}_{\textrm{ext}}$, in our case the dependence on $\dot{M}_{\textrm{ext}}$ is much weaker. For most of our simulations (with the exceptions of the low viscosity and high primary mass cases), the inner disc mass scales as $\dot{M}^{0.3}_{\textrm{ext}}$. Increasing $\dot{M}_{\textrm{ext}}$ leads to a faster migration rate which implies that the inner disc has less time to drain out, but a linear scaling between $M_{\textrm{d,in}}$ and $\dot{M}_{\textrm{ext}}$ is not necessarily expected.  Indeed, the dependence of the migration rate on the external accretion rate is non linear as it depends both on the surface density and the temperature at the inner edge of the outer disc, thus making it not straightforward to find a simple prescription to relate these two quantities.}

\begin{table}
\caption{Comparison of simulations results. We perform simulations for different values of ${\Mp =10^{6},\ 10^{7},\ 10^{8} \Msun}$,\ \  ${\dot{M}_{\mathrm{ext}}=0.1,\ 0.01 \Msun \uyr^{-1}}$,\ \ ${q=0.1,\  0.3}$ and ${\alpha=0.1,\ 0.01}$. In the Table, masses are in units of ${\Msun}$ and luminosities in units of ${L_{\mathrm{Edd}}}$. We report the inner disc mass at decoupling ${M_{\mathrm{d,in}}}$ and the luminosity peak just prior to merger ${L_{\mathrm{peak}}}$. For better comparison, we report the values obtained by \citet{2010MNRAS.407.2007C} in the last two columns, $M_{\mathrm{d,in}}^{(*)}$ and ${L_{\mathrm{peak}}^{(*)}}$. 
}
\label{table.results}
\begin{tabular}{*{6}{r}rr}
\hline
$\Mp$ & $q$ 	& $\dot{M}_{\mathrm{ext}}$ & $\alpha$ 	& $ M_{\mathrm{d,in}}$ & ${L_{\mathrm{peak}}}$  & $ M_{\mathrm{d,in}}^{(*)}$ &${L_{\mathrm{peak}}^{(*)}}$  \\
\hline
${10^6}$ &        &        &        &$\times {10^{-3}}$ &  &$\times {10^{-3}}$ &  \\
       &    0.1 &    0.1 &   0.1 &   0.95 &   1.97 & 0.0012 &  0.146    \\
       &        &        &   0.01 &     17 &   13.7 &  0.012 &  0.277  \\
       &        &   0.01 &   0.1 &   0.41 &   0.86 & 0.00012 &  0.015 \\
       &        &        &   0.01 &    8.4 &   6.44 & 0.0012 &  0.029  \\
       &    0.3 &    0.1 &   0.1 &   1.47 &    1.7 & 0.0001 &  0.038  \\
       &        &        &   0.01 &     25 &    8.8 &  0.001 &  0.023  \\
       &        &   0.01 &   0.1 &   0.68 &   0.92 & 0.000008 &  0.004 \\
       &        &        &   0.01 &     13 &   5.32 & 0.0001 &  0.002  \\
\hline
${  10^7}$ &        &        &        & $\times {10^{-3}}$ &        & $\times {10^{-3}}$ &      \\
       &    0.1 &    0.1 &   0.1 &    183 &   6.22 &   0.66 &  0.154 \\
       &        &        &   0.01 &   2970 &   20.8 &     46 &  1.077 \\
       &        &   0.01 &   0.1 &     83 &   1.38 &  0.073 &  0.018 \\
       &        &        &   0.01 &   1600 &   9.66 &   0.64 &  0.154 \\
       &    0.3 &    0.1 &   0.1 &    300 &    2.9 &  0.062 &  0.015 \\
       &        &        &   0.01 &    450 &   23.5 &   0.56 &  0.115 \\
       &        &   0.01 &   0.1 &    140 &   1.51 & 0.0062 &  0.001 \\
       &        &        &   0.01 &   2400 &   8.22 &  0.075 &  0.015 \\
\hline
${  10^8}$ &        &        &        &        &        &   &               \\
       &    0.1 &    0.1 &   0.1 &     19 &   2.67 &    0.2 &  0.5  \\
       &        &        &   0.01 &     34 &   0.47 &    0.9 &  2.154  \\
       &        &   0.01 &   0.1 &     11 &   1.31 &  0.029 &  0.069  \\
       &        &        &   0.01 &     35 &   0.48 &   0.18 &  0.438  \\
       &    0.3 &    0.1 &   0.1 &     26 &   2.15 &  0.024 &  0.052  \\
       &        &        &   0.01 &     34 &  0.189 &   0.15 &  0.323  \\
       &        &   0.01 &   0.1 &      22  &  1.24  & 0.0033 &  0.007  \\
       &        &        &   0.01 &     34 &   0.19 &  0.033 &  0.069  \\
\hline
\end{tabular}  
\end{table}

\subsection{Evolution of a fiducial run}
Let us describe the evolution of the system for a fiducial case with ${M_{\mathrm{p}}=10^{7}\Msun}$, ${q=0.1}$, ${\dot{M}_{\mathrm{ext}}=0.1\Msun \uyr^{-1}}$ and ${\alpha =0.1}$.

We choose to initialize the disc with the same choice of \citet{2010MNRAS.407.2007C}, i.e. with a low mass disc of mass ${10^{3}\Msun}$, appropriate for a Shakura-Sunyaev disc accreting at a very low rate, e.g. ${10^{-6}-10^{-7}}\Msun/$yr. Note that the purpose of this initial configuration of the disc is to give a simple realization of the planet-like regime in which the binary evolution takes place and it can be shown \citep[for a detailed discussion see][]{2010MNRAS.407.2007C} that the inner disc mass at decoupling -- and thus the peak luminosity just prior to merger -- does not depend on the initial mass of the disc.

\begin{figure*}
\centering
\includegraphics[width=0.49\textwidth]{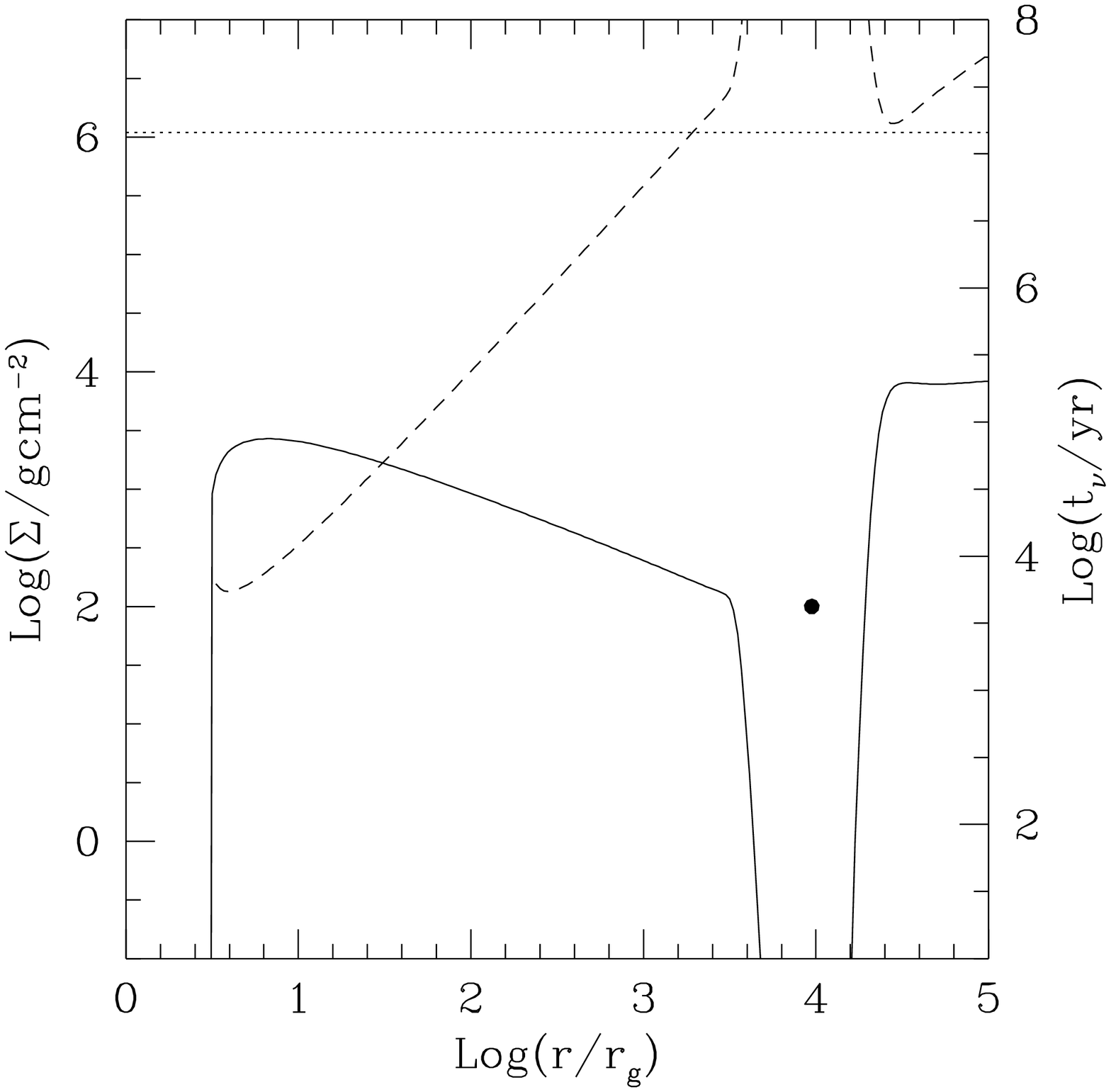}
\includegraphics[width=0.49\textwidth]{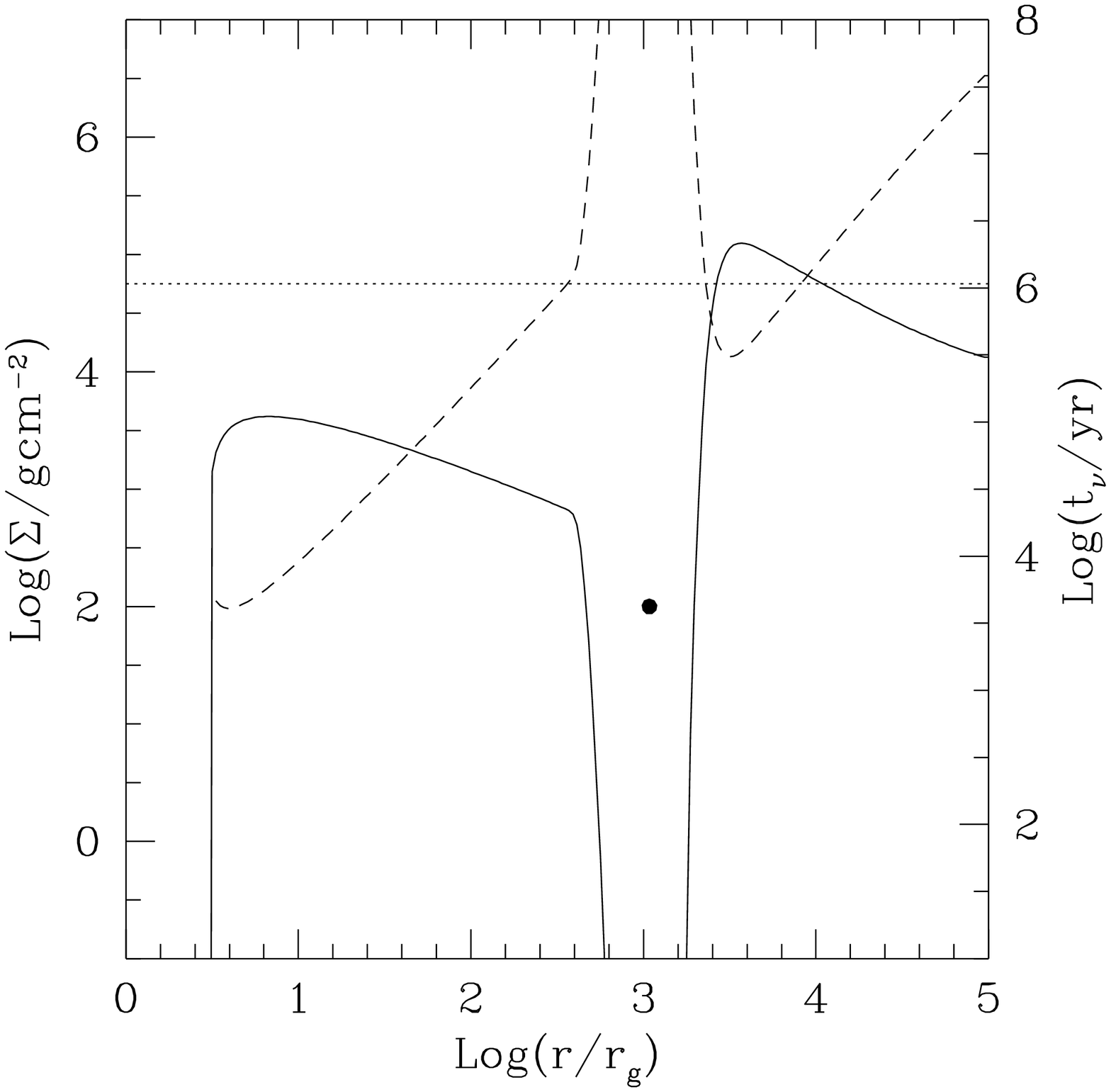}\\
\includegraphics[width=0.49\textwidth]{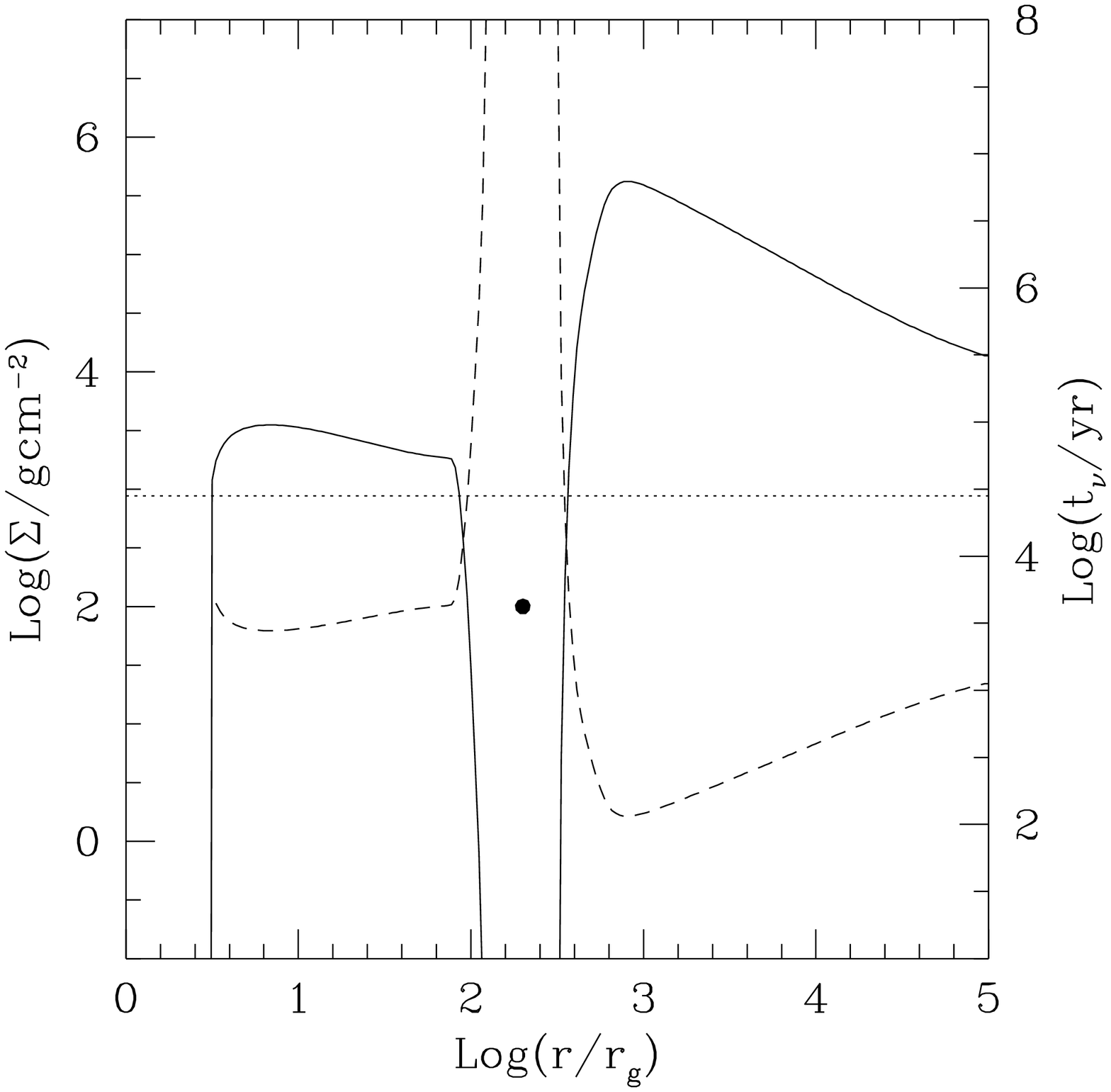}
\includegraphics[width=0.49\textwidth]{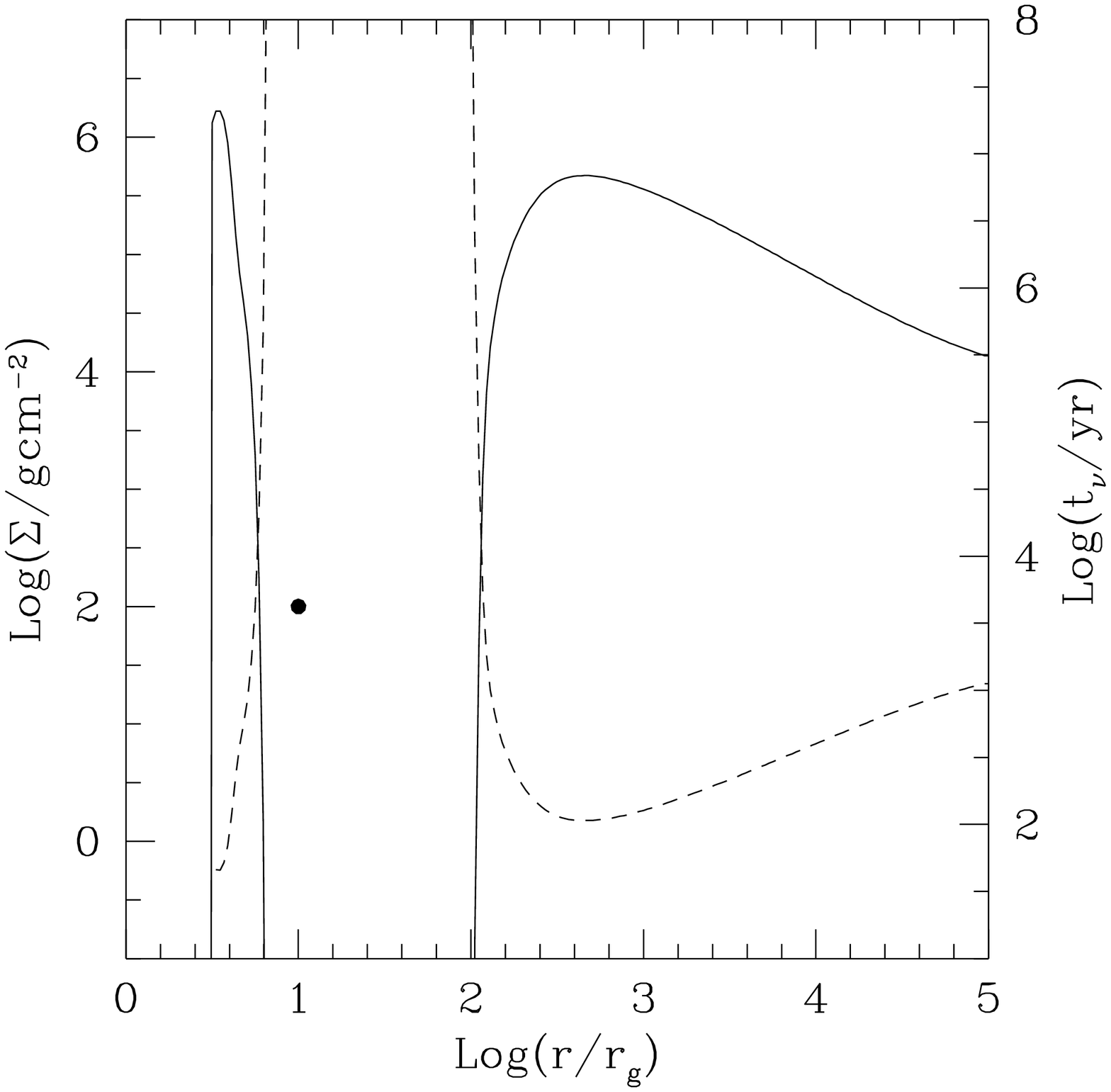}
\caption{Disc and binary evolution for the fiducial case with ${M_{\mathrm{p}}=10^{7}\Msun}$, ${q=0.1}$, ${\dot{M}_{\mathrm{ext}}=0.1\Msun \uyr^{-1}}$ and ${\alpha =0.1}$. Each snapshot shows the surface density profiles ${\Sigma}$ (solid lines), the viscous timescale profile ${t_{\nu}}$ (dashed lines, matched to the right axis),  the merger time ${t_{\mathrm{merge}}}$ (horizontal dotted line, matched to the right axis) and the secondary position (marked with a black dot). \mt{The snapshots have been taken respectively at 5 Myr (top-left), 12 Myr (top-right), 12.941 Myr (bottom-left) and 12.948 Myr (bottom-right) after the beginning of simulation.}}
\label{fig.snapshots.ourcode}
\end{figure*}

The general evolution of the system is very similar to the one described in \citet{2010MNRAS.407.2007C}. Following Figure \ref{fig.snapshots.ourcode}, in each snapshot we show the surface density profile (solid line), the viscous timescale (dashed line) and the merger timescale ${t_{\mathrm{merger}}=a/\dot{a}}$ (dotted line). Time evolution proceeds as follows: top-left, top-right, bottom-left, bottom-right. At the beginning (Figure \ref{fig.snapshots.ourcode}, top-left), the system reaches an initial quasi-stationary state in which the outer disc accumulates some mass due to the external accretion rate but the tidal torques are still not strong enough to push the secondary. 

Starting from this moment, the outer disc increased surface density begins to influence the secondary and to drive its inward migration. The gap edges are the locations where the viscous torques equal the tidal torques and, since the latter depend on the position of the secondary, the gap edges move together with the secondary. As a result the system self-regulates in such a way that the viscous timescale at the gap edges equals the merger timescale during the disc-driven phase \citep{2010MNRAS.407.2007C}. Then, as the disc-driven migration continues (Figure \ref{fig.snapshots.ourcode}, top-right), the gap moves inwards along with the secondary, as we expect for a Type II migration. The orbital radius of the secondary has decreased by one order of magnitude, and the inner disc has a slightly enhanced surface density. The viscous timescale at the gap edges continues to match the merger timescale (Figure \ref{fig.snapshots.ourcode}, bottom-left). When the binary separation has decreased by another order of magnitude (${a\approx 100 R_{\mathrm{S}}}$) the emission of GW rapidly takes over the tidal interaction and starts driving the migration. At this stage the outer disc is substantially decoupled from the evolution of the binary and the inner disc is pushed inwards more and more rapidly as the binary shrinks. Eventually (Figure \ref{fig.snapshots.ourcode}, bottom-right) the increasing velocity of the secondary (accelerated by the growing GW emission) forces the inner disc to accrete on a timescale much shorter than the viscous one, thus producing the spike in the surface density profile and the consequent flare in luminosity. In the last snapshot ${t_{\mathrm{merger}}}$ is not shown because it is much smaller (about days) than the viscous one.

\begin{figure}
\includegraphics[width=\columnwidth]{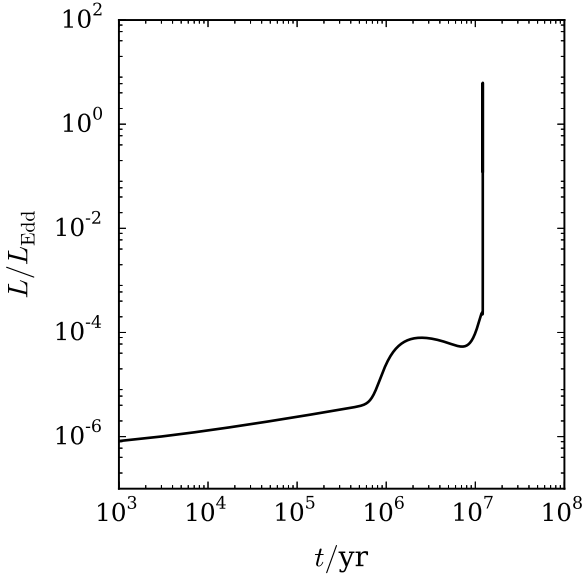}
\caption{Accretion luminosity on the primary BH as a function of time during the overall evolution of the disc-binary system \mt{for a fiducial run with ${\Mp=10^{7}\Msun,\ q=0.1,\ \dot{M}_{\textrm{ext}},\ \alpha=0.1}$}. The SMBHs coalesce around ${13}$ Myr, immediately after the apparent super-Eddington flare due to the rapid accretion of the inner disc.}
\label{fig.acc.lum}
\end{figure}

In Figure \ref{fig.acc.lum} we show the accretion luminosity on the primary BH as a function of time during the overall evolution of the simulation. 
\mt{During the first $10^6$ years the accretion rate is very low, and it reflects the initial condition for the disc surface density (see Section \ref{sec:results}). During this time the disc mass builds up as a consequence of the imposed external accretion rate, slowly increasing the disc luminosity to values closer to typical AGN values. As the disc mass grows, it pushes the secondary inwards, until, after roughly $10^7$ years, the secondary migration becomes dominated by gravitational wave emission.} Finally, it can be clearly seen the super-Eddington flare due to the extremely rapid accretion of the inner disc that occurs at the end of the GW-driven inspiral.

\section{Discussion}
\label{sec:discussion}

\subsection{Effects of tidal torque smoothing on luminosity peak}
Comparing our results with those by \citet{2010MNRAS.407.2007C} we see that these authors predict much lower inner disc masses and, consequently, much smaller peak luminosities for all the computed models by up to three orders of magnitude. For a better comparison, we report  their results in the last two columns of Table \ref{table.results}.
In order to explain the discrepancy in the estimate of the inner disc mass at the decoupling, we have performed several tests before finding that the issue was in the implementation of the torque. 

Before discussing the torque implementation, we would mention that a possible source of discrepancy may arise from the different viscosity calculation: as explained in Section \ref{sec.disc.energetics}, while we neglect radiation pressure when computing viscosity in order to maintain the disc stable, we take it into account when computing the disc thickness assuming that the disc is supported by both gas and radiation pressure. Therefore, differently from \citet{2010MNRAS.407.2007C} that completely neglect radiation pressure, our code is able to describe radiation pressure dominated regions as well. During our tests, we see that  the radiation pressure actually becomes dominant only in the inner disc in the very late phases just prior to merger and in the inner edge of the outer disc after a significant amount of gas has piled up. However, the estimate of the inner disc mass seems not to be affected by this choice. 

Another source of discrepancy might come from the way the gap is initially opened in the disc: while our code computes the gap opening consistently with the  torque calculated at each timestep, \citet{2010MNRAS.407.2007C} artificially open a cavity in the disc between  ${0.5a_{0}\leq R \leq 2 a_{0}}$ in order to simply model the gap clearing. At the very beginning of the simulation these different implementations produce visible differences in the gap width, however after a few Myr the gap width results determined by the intensity and the spatial distribution of the tidal torque rather than by the initial clearing.

\begin{figure*}
\centering
\includegraphics[width=0.49\textwidth]{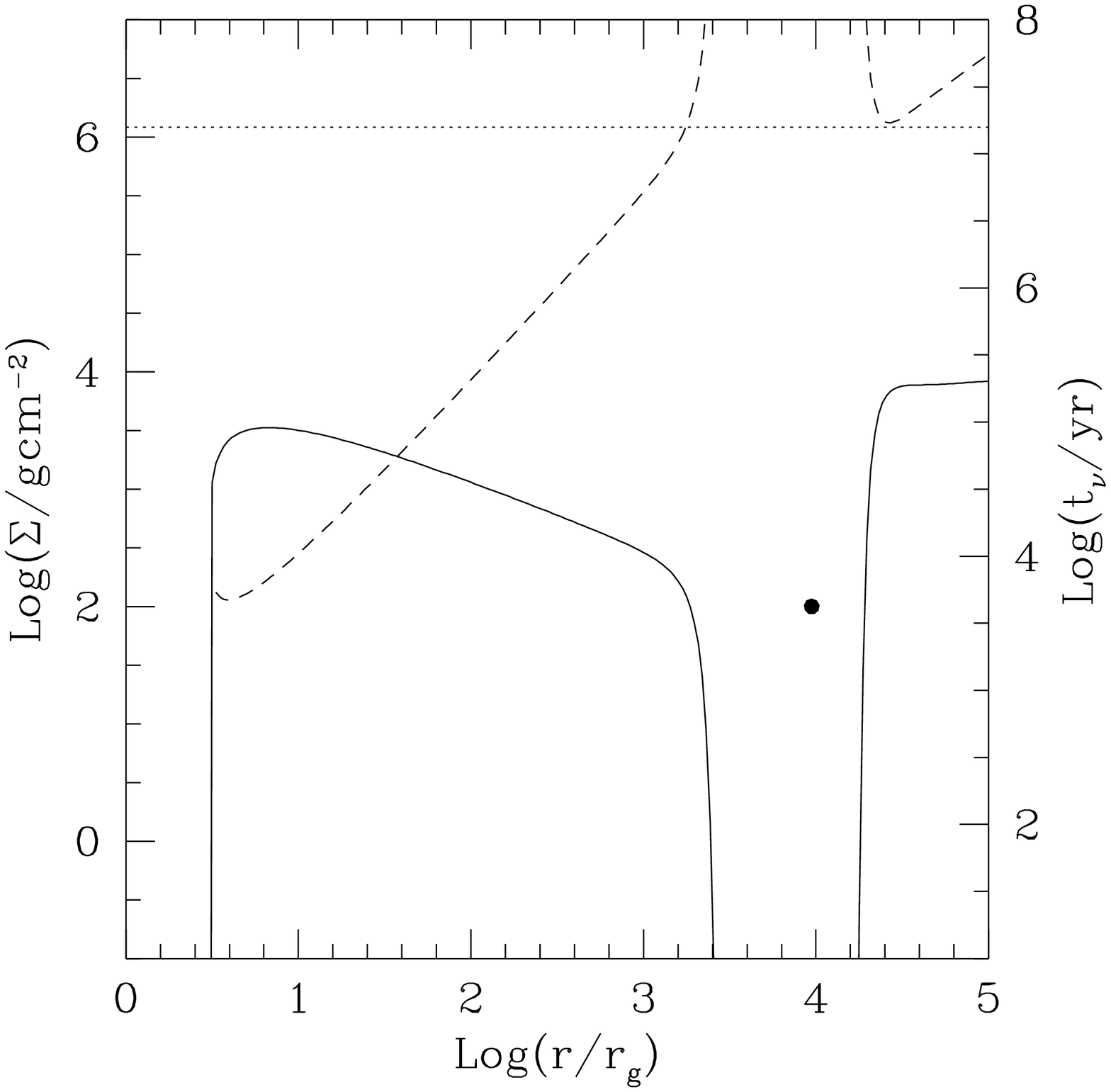}
\includegraphics[width=0.49\textwidth]{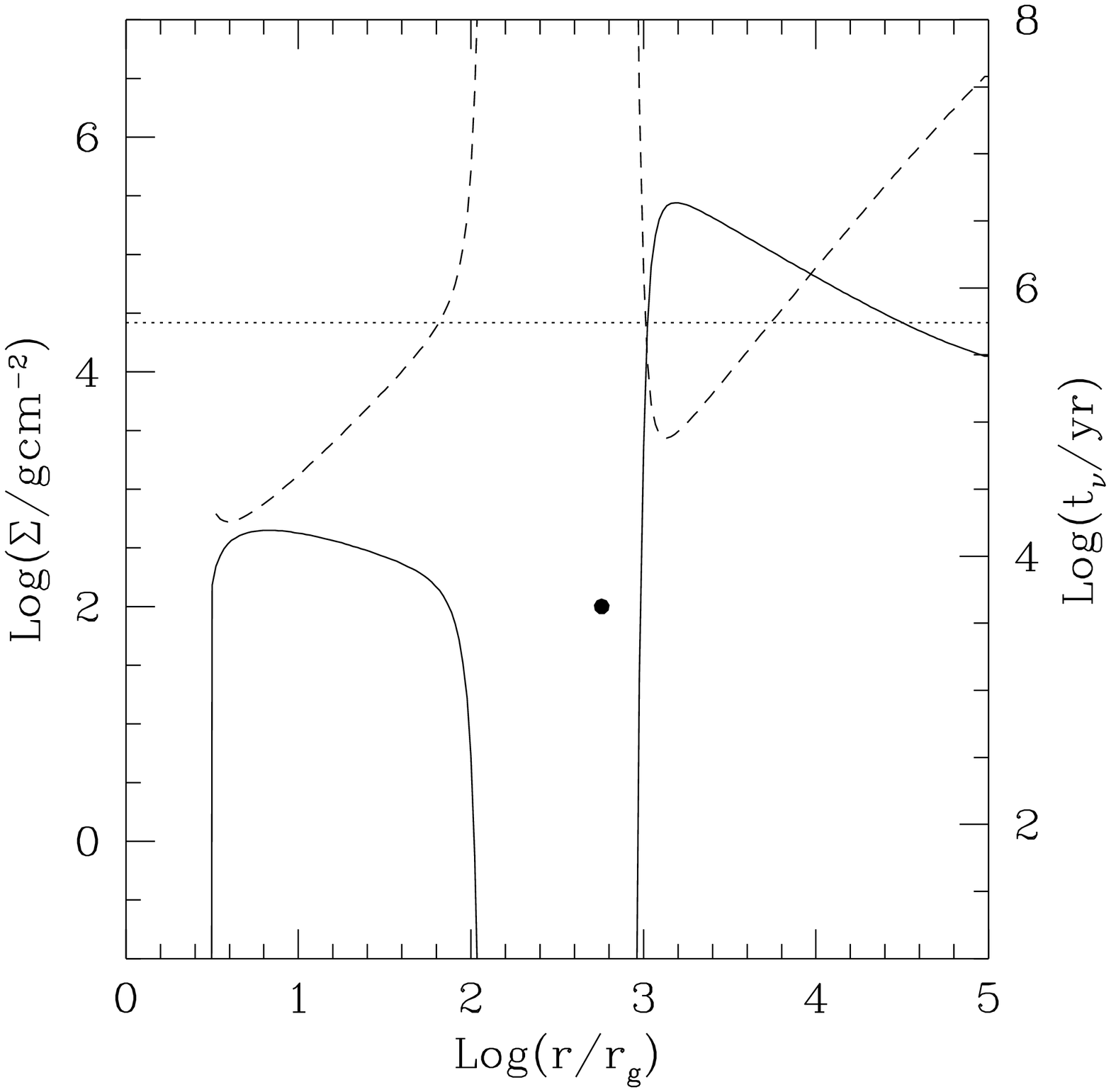}\\
\includegraphics[width=0.49\textwidth]{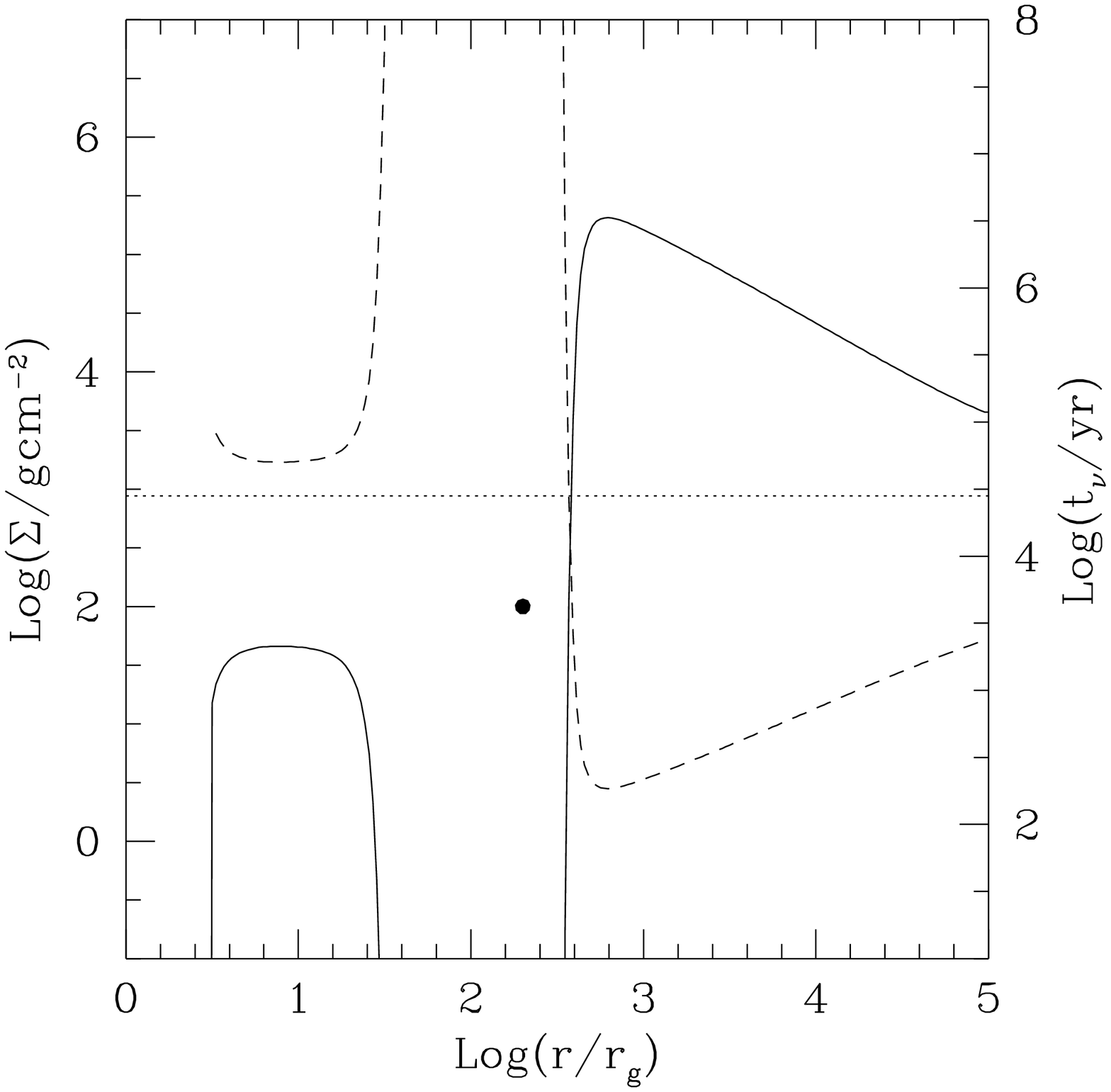}
\includegraphics[width=0.49\textwidth]{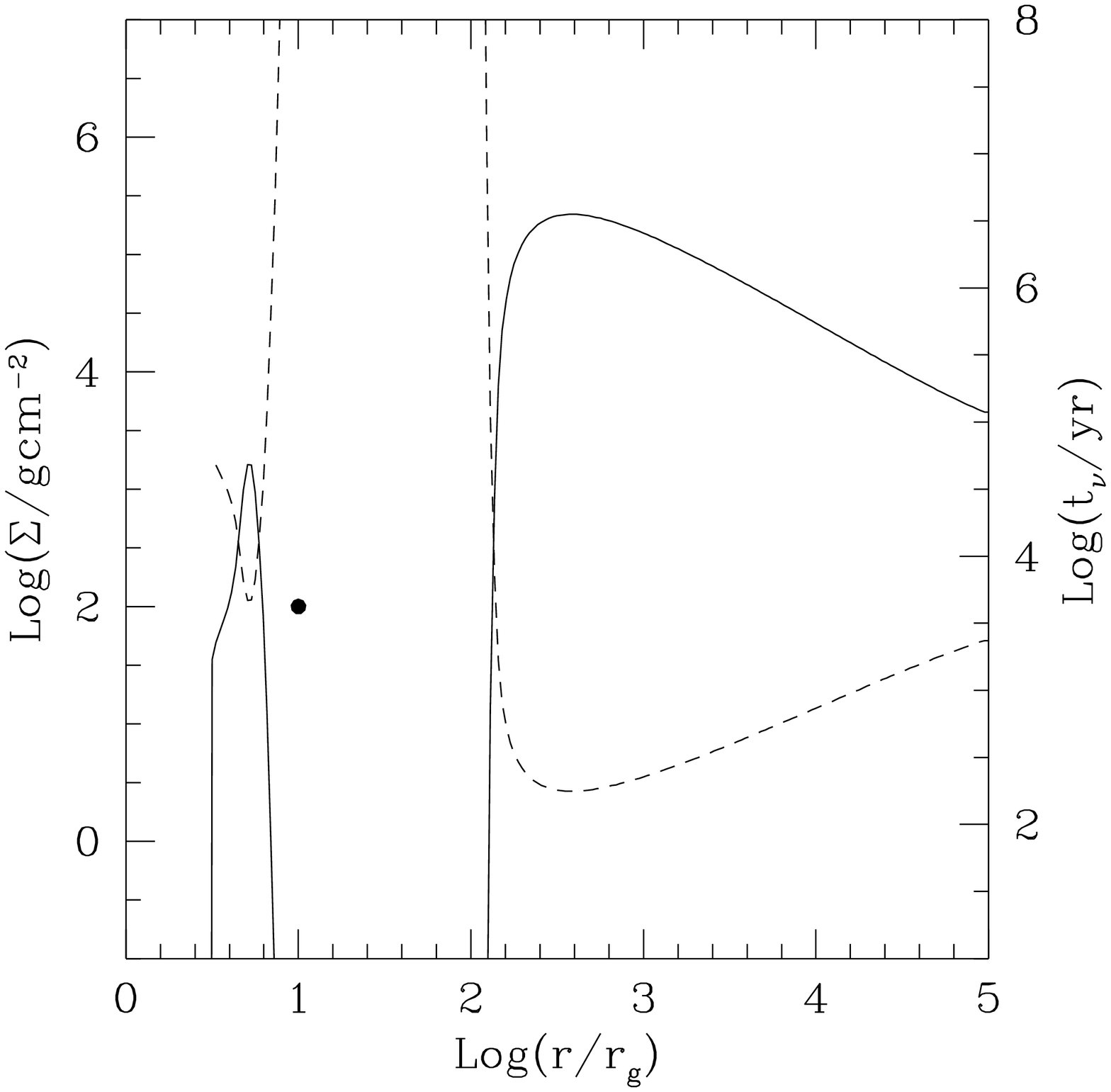}
\caption{Disc properties computed \textit{without} torque truncation for the fiducial case with ${M_{\mathrm{p}}=10^{7}\Msun}$, ${q=0.1}$, ${\dot{M}_{\mathrm{ext}}=0.1\Msun \uyr^{-1}}$ and ${\alpha =0.1}$. Each snapshot shows the surface density profiles ${\Sigma}$ (solid lines), the viscous timescale profile ${t_{\nu}}$ (dashed lines, matched to the right axis),  the merger time ${t_{\mathrm{merge}}}$ (horizontal dotted line, matched to the right axis) and the secondary position (marked with a black dot). The snapshots are taken at the same binary separations of Figure~\ref{fig.snapshots.ourcode}. Time evolution proceeds as follows: top-left, top-right, bottom-left, bottom-right.
}
\label{fig.snapshots.nosmoothing}
\end{figure*}

We thus conclude that the source of the discrepancy is in the implementation of the torque. Indeed, while for small mass ratios the LRs responsible for the gap formation are those closer to the satellite (and therefore the torque intensity is determined by the parameter ${f}$), for the larger mass ratios ${q \geq 0.1}$ that we are using here, the torque given by Eq. \eqref{eq.tidal.torque.original1} and \eqref{eq.tidal.torque.original2} is a significant overestimate, and thus needs to be smoothed out as we described in section \ref{sec.tidal.orque} above. 
In Fig. \ref{fig.snapshots.nosmoothing} we show the evolution of our fiducial model with ${M_{\mathrm{p}}=10^{7}\Msun}$, ${q=0.1}$, ${\dot{M}_{\mathrm{ext}}=0.1\Msun \uyr^{-1}}$ and ${\alpha =0.1}$, but now computed without smoothing the torque outside the Lindblad resonances. Note that the evolution shown here is almost identical to the one shown by \citet{2010MNRAS.407.2007C}. 

Comparing the snapshots in Figure \ref{fig.snapshots.nosmoothing} with those in Figure \ref{fig.snapshots.ourcode} it is apparent that, without smoothing the torque, the code produces a larger gap (the width being approximately ${0.94a}$ instead of ${0.7a}$) and smoother gap edges. Such wide gaps are an artifact of the 1D model with no smoothing and are not consistent with the more detailed 3D simulations by \citet{1994ApJ...421..651A}. As a result, starting from the initial stages (Figure \ref{fig.snapshots.nosmoothing}, \mt{top-left}) the inner disc is truncated at a much smaller radius (as we expect after the discussion in Section \ref{sec.tidal.orque}). As described in \citet{2010MNRAS.407.2007C} the surface density of the disc regulates in such a way to keep the viscous time at the outer edge equal to the migration time\footnote{\mt{Note that \citet{2014ApJ...792L..10D} and \citet{2014arXiv1411.3190D} have performed numerical simulations of Type II migration where the migration timescale did not exactly match the viscous timescale in the disc. This might be an interesting phenomenon but needs to be studied with greater detail, and in particular it is not clear whether such migration rates can be supported over long timescales, where a global balance of angular momentum could be established. A full discussion of this issue is clearly beyond 
the scope of the present paper.}}. If the outer edge of the inner disc is located much further in, this requires a much smaller surface density, thus leading to a significant underestimate of the inner disc mass. 

In this way, we can provide an analytic argument to quantify the extent to which the inner disc mass is underestimated if the torque is not smoothed. 
At the beginning of the simulation, when the binary separation is large (${a\approx 10^{4}R_{\mathrm{S}}}$), in the region where the inner disc is truncated by the tidal torques exerted by the satellite (at ${R\approx R_\mathrm{edge}}$), the disc is gas pressure dominated and opacity is dominated by Thomson scattering, and thus we can write viscosity as
\begin{equation}
\nu =\left(\frac{\alpha k_{\mathrm{B}}}{\mu m_{\mathrm{p}}} \right)^{4/3} \left(\frac{9}{16}\frac{\kappa_{\mathrm{T}}}{G \Mp\sigma_{\mathrm{SB}}} \right)^{1/3}\Sigma^{2/3}R\,,
\end{equation}
where we have substituted ${T_{c}}$ from Eq. \eqref{4.tcentral} in the ${\alpha-}$prescription \eqref{eq.alpha.prescription} and used ${H=c_{\mathrm{s}}/\Omega}$ that holds for a geometrically thin disc. With this expression for viscosity, we can estimate the viscous timescale at the inner gap edge:
\begin{equation}
t_{\nu}(R_{\mathrm{edge}})
\propto \frac{R_{\mathrm{edge}}^{2}}{\nu(R_{\mathrm{edge})}} \propto \frac{R_{\mathrm{edge}}}{\Sigma^{2/3}}\,.
\end{equation}
We have seen that during the secondary migration the gap moves together with secondary: this means that the viscous time at ${R_{\mathrm{edge}}}$ has to match the secondary migration time (Fig. \ref{fig.snapshots.nosmoothing}a,b,c). From the above equation we see that, if the inner disc is truncated at a smaller radius, then the surface density has to adjust itself to a lower value ${\Sigma(R_{\mathrm{edge}})\propto R_{\mathrm{edge}}^{3/2}}$, to give ${t_{\nu}(R_{\mathrm{edge}}) = t_{\mathrm{merger}}}$. To a first approximation we can then compute the inner disc mass:
\begin{equation}
M_{\mathrm{d,in}}=
\int_{R_{\mathrm{in}}}^{R_{\mathrm{edge}}}2\pi R \Sigma \d R\simeq
2\pi R_{\mathrm{edge}}^{7/2}\,,
\end{equation}
where the strong dependence on ${R_{\mathrm{edge}}}$ reflects the dependence of viscosity on surface density in the gas-pressure dominated region. From the above result we can now estimate \mt{the discrepancy in the inner disc mass computed by two different torque implementations. Adopting the same notation as in Figure \ref{fig.gap.edges}, ${R_{\mathrm{edge}}'}$ and ${R_{\mathrm{edge}}''}$ are the locations of the gap edges produced respectively by our code (with torque smoothing outside the LRs) and by the torque implementation used in \citet{2010MNRAS.407.2007C} (without torque smoothing). Hence we obtain that the ratio between the inner disc masses computed in the two cases is}:
\begin{equation}
\frac{M_{\mathrm{d,in}}'}{M_{\mathrm{d,in}}''}\propto \left(\frac{R_{\mathrm{edge}}'}{R_{\mathrm{edge}}''} \right)^{7/2}\,.
\end{equation}
For the fiducial case that we are considering, we get \mt{(at the time at which a gap has fully developed but the secondary has not yet started migrating)} ${R_{\mathrm{edge}}'\simeq 4.66\times 10^{3}r_{\mathrm{g}}}$ and ${R_{\mathrm{edge}}''\simeq 1.12\times 10^{3}r_{\mathrm{g}}}$ thus leading to:
\begin{equation}
\frac{M_{\mathrm{d,in}}'}{M_{\mathrm{d,in}}''}\approx 150\,,
\end{equation}
comparable to the discrepancy that we found between our results and those in \citet{2010MNRAS.407.2007C}. \mt{We have verified that this ratio is not strong function of time.}

\subsection{Limitations of the present calculations}
Some remarks should be made about the assumptions of the code, and the validity of the results. In our code the inner disc can only accrete on the primary BH, forced by the GW-driven inspiral of the secondary, thus neglecting any possible mechanism that is able to drain the inner disk. Therefore, the estimates of the peak luminosity that we report should be regarded as upper limits that hold in the optimistic case in which there is no inner disc leakage, e.g. across the gap or through jets. 

Some processes that we are not considering in our simulations might be responsible for a reduction of the inner disc mass in the late phases of the binary shrinking. Indeed, given the 1D nature of our code, by definition we are not able to describe off-plane or not cylindrically-symmetric gas flows. However these flows, that would require 2D or, better, 3D SPH simulations to be investigated,  might become relevant only in the innermost regions of the disc, where the tidal interaction and the accretion on the primary cause a significant heating of the gas, possibly allowing the disc to become geometrically thick. 

Another possible leakage of the inner disc may be due to a gas flow across the gap, as showed by \citet{2012MNRAS.423L..65B}. Gas flow across the gap has been studied by \citet{1999ApJ...514..344B,1996ApJ...467L..77A} and, more recently, by a variety of authors \citep{2012ApJ...749..118S, 2013MNRAS.436.2997D,2014ApJ...783..134F}.
In general, since the intensity of the torque scales with ${q^{2}}$, we expect that a low-mass secondary is less efficient in producing a clear gap, thus allowing a certain amount of gas to be funneled through the gap. As the secondary mass increases, the mass flow through the gap decreases significantly, reaching a minimum value of the order of 10 percent of the unperturbed accretion rate for $q\sim 0.1$ \citep{2013MNRAS.436.2997D}. 
For even larger secondary masses, the cavity is significantly deformed, becoming elliptical and allowing gas to enter the cavity. Thus the mass flow through the gap increases with increasing secondary mass for $q \ga 0.1$. 
For the range of mass ratios that we are interested in here ($q\sim 0.1-0.3$) 
the leakage from the outer to the inner disc might be as high as 40 percent of the unperturbed accretion rate, further enhancing the fossil disc mass \citep{2013MNRAS.436.2997D}. 
Note, however, that in most cases these simulations, for numerical convenience, adopt very large values for the disc aspect ratio ($H/R\sim 0.1$), much larger than the values obtained here ($H/R\sim 0.001$). 

Recent 3D magneto-hydrodynamic (MHD) simulations by \citet{2012ApJ...749..118S} suggest another mechanism that might favor an increase in the inner disc mass through the gap. Indeed, they show that in a highly ionized gas disc, where the magneto-rotational instability (MRI) can effectively induce angular momentum transport through MHD turbulence, the gas flow across the gap results enhanced by more than one order of magnitude with respect to calculations that neglect the coupling between the gas and the magnetic fields.

\mt{Finally, another limitation of the model comes from the fact that the formulation of the problem as adopted here is strictly valid only if the mass ratio $q\ll 1$. For larger mass ratios (possibly already for the ${q=0.1}$ case discussed here) both the primary and the secondary will rotate around the common center of mass with an angular velocity that takes into account both the primary and secondary mass. We have kept this simple formalism for an easy comparison with the previous work by \citet{2010MNRAS.407.2007C}, but we note here that the detailed results might be somewhat modified in 2D or 3D, especially for the largest values of ${q}$.}

\section{Conclusions and outlook}
\label{sec:conclusions}
In this work we have revisited the problem of estimating the fossil disc mass just prior to a SMBH merger in order to explain the discrepancy among the contradictory results obtained by different authors. While \citet{2002ApJ...567L...9A} and \citet{2009MNRAS.398.1392L} predict large values of the inner disc mass at decoupling (that thus result in powerful super-Eddington flares), \citet{2010MNRAS.407.2007C} find values for the inner disc mass smaller by several orders of magnitude (that therefore produce significantly less powerful flares). These previous works consider very different scenarios for the disc-binary system and a direct comparison of the estimates of inner disc mass is not trivial. 

In order to find the origin of the discrepancy, we have implemented a 1D hydrodynamic code that solves the coupled evolution of the disc and the binary in a self-consistent way. With our code we have been able to reproduce the results of \citet{2002ApJ...567L...9A} and \citet{2009MNRAS.398.1392L} with an extremely good agreement. Then, we executed the simulations performed by \citet{2010MNRAS.407.2007C} adopting their same initial configurations and parameters: we found inner disc masses larger by up to three orders of magnitude than those obtained by \citet{2010MNRAS.407.2007C}. As discussed in Section \ref{sec:discussion}, we found that the origin of the discrepancy lies in the fact that for mass ratios larger than ${0.1}$ the usual analytic formula for the tidal torque provides an unphysical contribution at large distances from the satellite, resulting in an artificially wide gap, and therefore needs to be smoothed out. In our simulations, we smooth the torque in such a way to reproduce the correct gap sizes as found in 3D simulations by \citet{1994ApJ...421..651A}. Differently from us, \citet{2010MNRAS.407.2007C} do not perform the torque smoothing  (private communication with P. Chang) and thus obtain systematically reduced fossil disc masses and peak luminosity.

It is worth remembering that these 1D computations neglect gas streams that depart off the plane of the disc and that would require a more realistic 3D treatment. 
These effects might enhance or decrease the actual luminosity during the merger. On the one hand, for a given mass of the fossil disc, the analysis of \citet{2012MNRAS.423L..65B} shows that up to 80 percent of the inner disc mass might leak out of the secondary orbit and escape accretion during the gravitational wave dominated inspiral. However, the limitations of such 2D computations and the unclear dependence of this kind of leakage on the disc-binary parameters require further investigations. On the other hand, leakage from the outer to the inner disc during the earlier disc-dominated phase would have the opposite effect of increasing the fossil disc mass above what estimated here. Indeed, it has been shown that MRI \citep{2012ApJ...749..118S} and binary eccentricity \citep{1996ApJ...467L..77A} are effective in refilling the inner disc through the gap, possibly ending with a further enhancement of the accretion rate on the primary BH. 

As a conclusion, taking our estimates as a lower limit to the fossil disc mass, even if, as obtained by \citet{2012MNRAS.423L..65B} the inner disc is depleted by around 80 percent, we would still have a super-Eddington flare in most cases, confirming that SMBH mergers are particularly promising for observational purposes.

\section*{Acknowledgements}
We thank Philip Chang for useful discussions and the anonymous referee for constructive report. MT thanks Giovanni Rosotti for valuable discussions about the numerical and the coding aspects of this work and  Agnese Miani for the precious support during the writing of this paper.
MT acknowledges the support by the DFG cluster of excellence
Origin and Structure of the Universe (\href{http://www.universe-cluster.de}{www.universe-cluster.de}).

\bibliography{mt_bib}

\bsp
\label{lastpage}
\end{document}